# Non-saturating magnetoresistance in heavily disordered semiconductors

M. M. Parish* & P. B. Littlewood* †

*Cavendish Laboratory, University of Cambridge, Cambridge, CB3 0HE, UK

†National High Magnetic Field Laboratory, Pulsed Field Facility, LANL, Los Alamos NM 87545

**The resistance of a homogeneous semiconductor increases quadratically with magnetic field $H$ at low fields and, except in very special cases, saturates at fields $H$ much larger than the inverse of the carrier mobility, a number typically of order 1 Tesla[1,2]. Here, we argue that a macroscopically disordered and strongly inhomogeneous semiconductor will instead show a non-saturating magnetoresistance, with typically a quasi-linear behaviour $\Delta R \equiv R(H) - R(0) \propto H$ up to very large fields, and possibly also extending down to very low fields, depending on the degree of inhomogeneity. We offer this as a possible explanation of the observed anomalously large magnetoresistance in doped silver chalcogenides[3-5]. Furthermore, our model of an inhomogeneous semiconductor can be developed into magnetoresistive devices that possess a large, controllable, linear response.**

Recently, an anomalously large magnetoresistance was observed in two doped silver chalcogenides, $Ag_{2+\delta}Se$ and $Ag_{2+\delta}Te$, where the resistance displayed a positive linear dependence on the magnetic field over the temperature range 4.5K to 300K, without any signs of saturation at fields as high as 60T (refs 3, 5). These characteristics make the compounds ideally suited for the development of magnetoresistive devices such as magnetic field sensors[5], but the origin of the linear magnetoresistance still remains unclear. The silver chalcogenides are narrow-gap semiconductors[6], so conventional theories predict that the magnetoresistance should saturate at large fields, unlike what is observed. Moreover, the silver chalcogenides possess no magnetic moments, therefore the magnetoresistance cannot be spin-mediated like the colossal magnetoresistance of the manganites[7]. Polycrystalline metals may also exhibit a linear magnetoresistance[8], but this behaviour requires the presence of open Fermi surfaces, which is not the case here. Currently, the only proposed explanation for the silver chalcogenide phenomenon is 'quantum magnetoresistance' of Abrikosov[9]. However, doped silver chalcogenides are granular materials[4] and a linear magnetoresistance has also been observed in metals with surface imperfections[10, 11] and in disordered indium antimonide[12]. Therefore, an alternative hypothesis is that the linear magnetoresistance of the silver chalcogenides results from large spatial fluctuations in the conductivity of the material, due to the inhomogeneous distribution of silver ions. A major advantage of such a classical theory is that the qualitative behaviour of the magnetoresistance is not sensitive to microscopic processes and, thus, only depends on temperature via changes in the macroscopic conductivity, which is consistent with experiment[3].

There have been extensive theoretical investigations into the conductivity of classical, inhomogeneous media[13], but the majority of them focus on the zero magnetic field case. The only known situation that yields a linear magnetoresistance is that of an isotropic medium with a low volume fraction $c$ of insulating inclusions[14, 15], but linearity only holds for exceedingly high fields $\beta c \gg 1$ (where $\beta = \mu H$, $\mu$ is the carrier mobility and $c \ll 1$), whereas the silver chalcogenide's magnetoresistance continues to be linear down to fields as low as 1Oe (ref. 3). Whilst an effective medium approximation gives the same result for a higher volume fraction of insulating inclusions, it still only applies for high fields $\beta \gg 1$ (ref. 16). Systems with continuously varying fluctuations in the conductivity have been shown to have a non-saturating magnetoresistance in the weak disorder limit[17,18], but here $\Delta R \propto \beta^{2/3}$ and the calculations are once again only relevant at extremely high magnetic fields $\beta\Delta >> 1$, where the disorder width $\Delta << 1$. Thus, the main limitation of the current literature is that it only deals with weak disorder, so high magnetic fields are then required to obtain a solution with nontrivial behaviour.

Here, we model a strongly inhomogeneous conductor problem by discretization into a random resistor network which we analyse numerically. Standard networks constructed from 2-terminal resistors are inadequate for describing systems in magnetic fields since there are not enough terminals to take account of the Hall components, so $\Delta R = 0$. The simplest discrete model of magnetoresistance is a two-dimensional square lattice constructed of 4-terminal resistors, with an external magnetic field applied perpendicular to the network. The advantage of such a network is that its sharp interfaces put it in a different class to continuous media, so that large conductivity fluctuations are already built into the system even when all the resistors are identical. Note that this network can only produce a transverse magnetoresistance, not a longitudinal one, as this requires a three-dimensional network.

We will consider the resistor unit to be a homogeneous disk with 4 current terminals and 4 voltage differences between terminals, as shown in Fig. 1(a). These currents $\iota$ and voltages $v$ are connected via a $4\times 4$ matrix $z$:

$$v_i = z_{ij}\iota_j \qquad (1)$$

The matrix elements $z_{ij}$ can be determined by solving the Laplace equation for the electric potential of a homogeneous, conducting disk, using the currents as boundary conditions, and they are found to have the form:

$$z_{ij} = \frac{\rho}{2\pi\varphi\tau}(c_{ij} + d_{ij}\beta) \qquad (2)$$

where $\rho$ is the resistivity, $\tau$ is the disk thickness, and $\varphi$ is the angular width of each terminal, whilst $c_{ij}$ and $d_{ij}$ are constants dependent on the geometry. A random resistor network can thus be created by varying any of the parameters present in Eqn.(2), although in practice we only need to consider two quantities: $r = \rho/2\pi\varphi\tau$ and $\mu$. (To ensure that the voltages $v$ are well defined at high magnetic field, we should require that the terminals consist of ideal thin metal contacts, $\varphi << 1/\beta$. )

We calculate the magnetoresistance of the network by using Kirchoff's laws to determine the network impedance matrix $Z$ that relates the input currents to the voltages at the input terminals, that is $V = ZI$. As depicted in Fig. 1(b), we consider the input currents to be non-zero only on two parallel edges of the network and we also ground one of these terminals to provide a point of reference for the voltages and impose current conservation. Thus, a network of $N \times M$ resistors will possess $2N - 1$ input currents and $2N - 1$ input voltages. If we now apply a constant potential difference $U$ across the network, the effective resistance $R_{NM}$ of an $N \times M$ network is given by

$$R_{NM} = \frac{U}{\sum_i I_i} = \frac{U}{\sum_i Z_{ij}^{-1}V_j} \qquad (3)$$

where the sum over input currents is performed along one edge. In the limit where $N \to \infty$ for a square network, $R_{NN}$ will give us the magnetoresistance of a strongly inhomogeneous medium.

Whilst Eqn.(3) is awkward to solve analytically for large networks, some general properties of the magnetoresistance can be extracted from the symmetries of $Z$. It can be demonstrated that

$$Z = S + \beta A \qquad (4)$$

where $S$ and $A$ are symmetric and antisymmetric matrices, respectively, which are independent of $\beta$ in both the limits $\beta \to 0$ and $\beta \to \infty$. This result is not surprising since the matrix $z$ in Eqn. (2) also has this form when one current terminal has been grounded. Indeed, this structure can be justified on physical grounds since as antisymmetric matrix implies dissipationless current flow, which is consistent with motion in a magnetic field. Therefore, $Z$ becomes antisymmetric as $\beta \to \infty$. Moreover, it will be an odd antisymmetric matrix since $Z$ is always a $(2N-1) \times (2N-1)$ matrix, so an eigenvalue $\lambda_0$ will approach zero at large fields. If we explicitly take out the factor $\beta$ so that $Z = \beta Z_\beta$ we can write the sum of input currents in Eqn. (3) as

$$\sum_i I_i = \frac{1}{\beta} \sum_i \sum_n \frac{w_{n,i} w_{n,j}^T}{\lambda_n} V_j \qquad (5)$$

where $w_n$ and $\lambda_n$ are the nth eigenvector and eigenvalue of $Z_\beta$, respectively. Due to the zero eigenvalue in the denominator, the zeroth term will determine whether the current goes to zero as $\beta \to \infty$ since all other terms vanish in this limit.

For the simplest case where all resistors in the network are identical, we obtain

$$\sum_i I_i \propto \begin{cases} \dfrac{U\delta_N(\beta)}{2N-1}, & \text{if } N \text{ is even} \\[2ex] \dfrac{U}{2N-1}, & \text{if } N \text{ is odd} \end{cases} \qquad (6)$$

where $\delta_N(\beta)$ is a high-field correction factor that vanishes as $\beta \to \infty$. Hence, $N \times M$ networks with even $N$ will exhibit a non-saturating magnetoresistance, whilst networks with odd $N$ will exhibit saturation except when the network is infinite.

A numerical analysis of $N \times N$ uniform networks produces magnetoresistances which all agree with Eqn. (6). Plotted in Fig. 2(a) is the normalised magnetoresistance $\Delta R(H)/R(0)$ as a function of magnetic field for odd-$N$ square networks of different sizes. The saturation level scales linearly with $N$ as predicted by Eqn. (6), and $\Delta R(H)/R(0) \propto \beta^2$ when $\beta < 1$ since $\Delta R$ must always be an even function of $H$ due to symmetry. Most importantly, we can see that the magnetoresistance curves collapse onto a straight line as $N \to \infty$ so that $\Delta R(H)/R(0) \propto \beta$ when $\beta > 1$ for infinite networks. The magnetoresistances of even-$N$ networks also collapse onto a linear curve in the infinite size limit, as shown in Fig. 2(b), but the limit is approached from above instead of below. Moreover, finite even-$N$ networks of disks always exhibit a non-saturating magnetoresistance, unlike the single, homogeneous, van der Pauw disk with an embedded concentric inhomogeneity, which exhibits an extremely large but saturating magnetoresistance[19, 20]. This non-saturating, linear behaviour makes large networks ideal candidates for sensors of large magnetic fields.

For infinite networks, $\Delta R(H)/R(0) \propto \mu$ when $\beta > 1$ and it is independent of charge carrier density unlike in Abrikosov's theory, so this explains why $\Delta R(H)/R(0)$ of the silver chalcogenides decreases with increasing temperature as this corresponds to a decrease in $\mu$ due to phonon excitations. In addition, it is consistent with other experiments which indicate that $\mu$ is the most important optimisation variable[21]. However, the crossover from linear to quadratic behaviour occurs at $H \sim \mu^{-1}$, which is much higher than the crossover field of the silver chalcogenides.

Of course, inhomogeneous materials are generally random in nature, so it is necessary to consider random resistor networks and see whether they fundamentally change the uniform network results. In eqn. (4), a random resistor network corresponds to a random matrix $A$, and this situation is more complex than the uniform case since the zero eigenvalue is associated with a distribution of eigenvectors when $N$ is large. Nonetheless, we would also expect the magnetoresistance to be non-saturating as $N \to \infty$ since the distribution of eigenvector elements $w_{0i}$ will be centred on zero so summing over these eigenvectors in Eqn. (5) gives zero.

Solving eqn. (3) numerically, we find that for finite random networks, $\Delta R$ depends on the particular network configuration so there is a large range in the behaviour, with certain configurations giving $\Delta R \propto H$ for networks as small as $4 \times 4$. For large networks, a visualisation of the current paths in the network helps to motivate the physics. As shown in Fig. 3, the currents behave in a counterintuitive manner at high magnetic fields since some currents reverse direction, creating loops within the inhomogeneous system. The voltage landscape is also nontrivial, although we can see that the major current paths almost exactly follow the voltage contours, as expected at high fields. An interesting observation is that the current paths are perpendicular to the applied voltage a significant proportion of the time. This leads us to expect a linear magnetoresistance for the network, since we would expect current flowing perpendicular to the longitudinal voltage to contribute the Hall resistance $R_h \propto H$ to the effective magnetoresistance.

As network size increases, the range in behaviour diminishes and the proportion of network configurations giving a nonsaturating $\Delta R$ increases in agreement with the analytical prediction. Furthermore, the magnetoresistance averaged over many network configurations gains a more linear dependence on magnetic field for large enough network sizes, so this coupled with the decreasing range in behaviour implies that $\Delta R \propto H$ for infinite random networks. Fig. 4 depicts the linear, averaged $\Delta R(H)/R(0)$ of $20 \times 20$ networks for different mobility distributions which are all assumed to be Gaussian. Note that we can include positive charge carriers (holes) in this model by allowing the mobility to be positive as well as negative. At sufficiently large magnetic fields we see that $\Delta R(H)/R(0) \propto \langle \mu \rangle$ for $\Delta \mu / \langle \mu \rangle < 1$ and $\Delta R(H)/R(0) \propto \Delta \mu$ for $\Delta \mu / \langle \mu \rangle > 1$, where $\langle \mu \rangle$ is the average mobility and $\Delta \mu$ is the width of the mobility disorder, so the magnetoresistive response is also strongly controlled by $\mu$ in the random system. In particular, the crossover field is $\langle \mu \rangle^{-1}$ for $\Delta \mu / \langle \mu \rangle < 1$ and $(\Delta \mu)^{-1}$ for $\Delta \mu / \langle \mu \rangle > 1$. Therefore, even when the characteristic field $\langle \mu \rangle^{-1}$ is of order 1T, the measured crossover field of a disordered semiconductor can be several orders of magnitude smaller, provided the mobility disorder is large.  In particular, we propose that the large, linear magnetoresistance observed in silver telluride when $\langle \mu \rangle \cong 0$ (ref. 4) is due to strong disorder in the mobility.

The silver chalcogenides are promising candidates for magnetic field sensors, but the size scale of the composition fluctuations is greater than 100nm (the bound is from unpublished small-angle-neutron-scattering data taken by M-L Saboungi, C Glinka and TF Rosenbaum).  This could limit their application to use as macroscopic sensors rather than nanoscopic sensors, such as read-heads, even if their sensitivity could be enhanced.  However, one can, in principle, fabricate arrays of miniature disks onto any conducting material, and thus develop magnetic field sensors that are cheap to manufacture and that have responses controlled by the material mobility.

Further work is required to extend the random resistor network model to 3 dimensions and to obtain the exact magnetic field dependence of the magnetoresistance for the infinite random network using, for example, random matrix theory, but this model is otherwise very successful in explaining the anomalous magnetoresistance observed in silver chalcogenides.  The networks display a non-saturating magnetoresistance which is a desirable property of magnetic field sensors, they can possess a linear magnetoresistance from low to high magnetic fields like the silver chalcogenides, and the linearity at low fields increases with growing disorder.

This work was supported by the Association of Commonwealth Universities and the Cambridge Commonwealth Trust. PBL thanks the NHMFL for hospitality during the final drafting of this article. The NHMFL is supported by the National Science Foundation, the state of Florida and the US Department of Energy.

Correspondence should be addressed to M.M.P. (e-mail: mmp24@phy.cam.ac.uk).

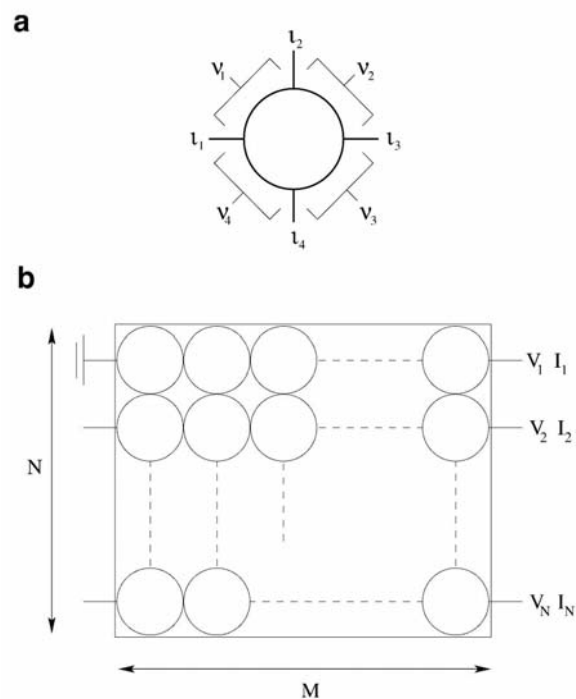

**Figure 1** 4-terminal network resistor unit and a schematic diagram of an $N \times M$ resistor network. **a**, The resistor unit is a homogeneous, conducting disk with currents $\iota$ entering the terminals and voltage differences $v$ between the terminals. **b**, The $N \times M$ resistor network has voltage $V_i$ and current $I_i$ associated with the $i$-th input terminal on one of the vertical edges, and the current is zero on the horizontal edges. One terminal is grounded to provide a reference for the voltages, and the current at this grounded terminal is ignored because of current conservation. In order to model the measured

magnetoresistance $\Delta R$ of a macroscopic sample, a constant potential difference across the network is applied by completely grounding the left vertical edge and setting the right vertical edge to a constant voltage $U$.

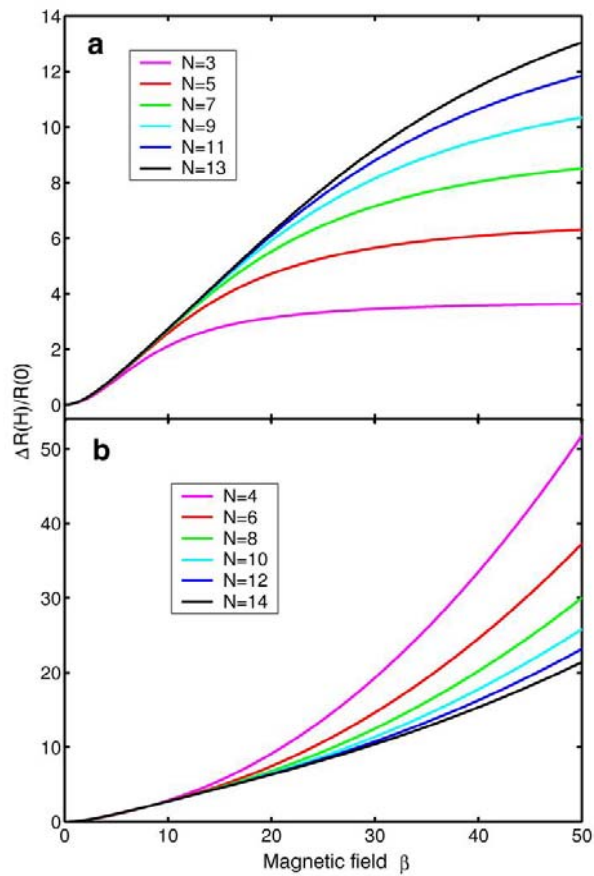

**Figure 2** Normalised magnetoresistance $\Delta R(H)/R(0)$ as a function of dimensionless magnetic field $\beta$ for different sized $N \times N$ uniform networks. **a,** For networks of odd $N$, the magnetoresistance saturates at high fields at a level that scales linearly with $N$, while at low fields, it crosses over from a linear to a quadratic dependence. As $N \to \infty$, the magnetoresistance curves collapse onto a straight line. **b,** Networks of even $N$ always exhibit a non-saturating magnetoresistance, which collapses onto a straight line from above as $N \to \infty$.

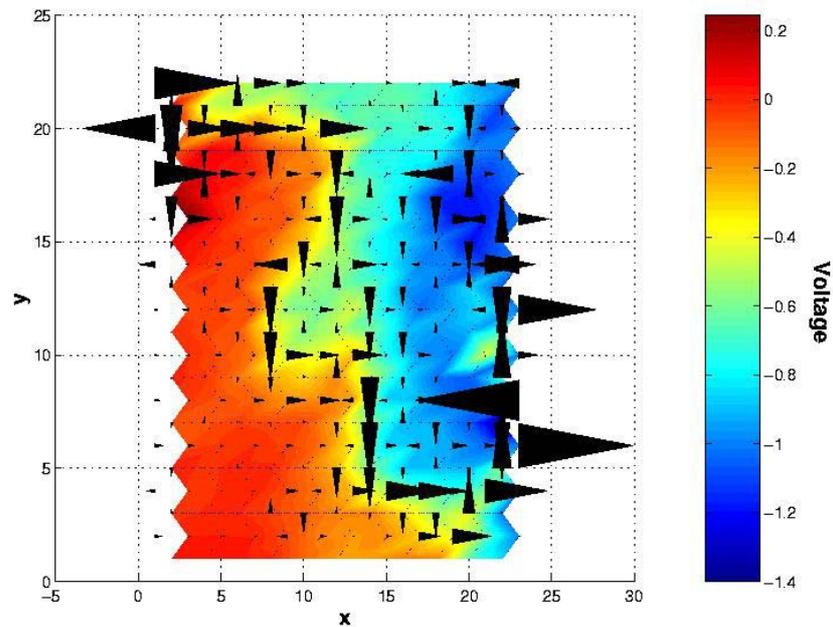

**Figure 3** Visualisation of currents and voltages at large magnetic field in a $10 \times 10$ random network of disks with radii 1 (arbitrary units), where the potential difference $U = -1\text{V}$. The black arrows represent the currents, where arrow size depicts the magnitude of the current. The major current path is perpendicular to the applied voltage a significant proportion of the time, which implies that the magnetoresistance is provided internally by the Hall effect, which is therefore linear in $H$.

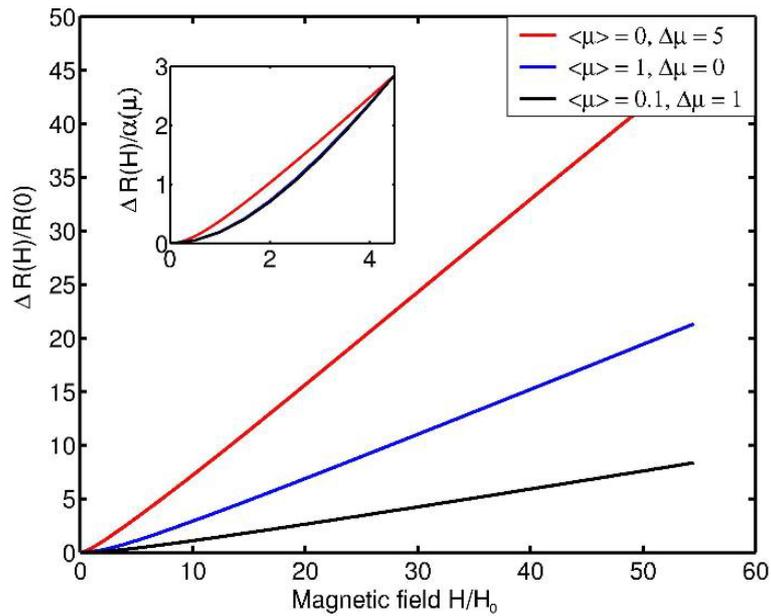

**Figure 4** Average normalised magnetoresistance $\Delta R(H)/R(0)$ as a function of dimensionless magnetic field $H/H_0$ of $20 \times 20$ random resistor networks for different mobility distributions, where $H_0 = 1\,\text{kOe}$ is a typical field scale. The magnetoresistance was averaged over 10 random network configurations and the mobility distributions were taken to be Gaussian and measured in units of $H_0^{-1}$. Strong mobility disorder results in a large magnetoresistive response. Inset: By scaling the curves so that they all have the same magnetoresistance at around 4kOe, it can be seen that linearity continues down to lower fields when the mobility disorder is large.